\documentclass[twocolumn,showpacs,preprintnumbers,amsmath,amssymb,prl]{revtex4}

\usepackage{graphicx}
\usepackage{dcolumn}
\usepackage{bm}
\usepackage{tabularx}
\usepackage{amsmath}

\begin{document}

\title{Reply to comment by T. Terashima {\it et al.} on ``Quantum criticality and nodal superconductivity in the
FeAs-based superconductor KFe$_2$As$_2$"}

\author{J. K. Dong,$^1$ S. Y. Zhou,$^1$ T. Y. Guan,$^1$ H. Zhang,$^1$ Y. F. Dai,$^1$ X. Qiu,$^1$ X. F. Wang,$^2$ Y. He,$^2$ X. H. Chen,$^2$ S. Y. Li$^{1,*}$}

\affiliation{$^1$Department of Physics, Surface Physics Laboratory (National Key Laboratory), and Laboratory of Advanced Materials, Fudan University, Shanghai 200433, China\\
$^2$Hefei National Laboratory for Physical Science at Microscale and
Department of Physics, University of Science and Technology of
China, Hefei, Anhui 230026, China}

\date{\today}


\maketitle

In our recent Letter \cite{JKDong}, we report the demonstration of a
field-induced antiferromagnetic quantum critical point (QCP) and
nodal superconductivity in KFe$_2$As$_2$. The evidences for a QCP
include non-Fermi-liquid $\rho(T) \sim T^{1.5}$ at the upper
critical field $H_{c2}$ = 5 T and the development of a Fermi liquid
state with $\rho(T) \sim T^2$ when further increasing the field. The
coefficient $A$ of the $T^2$ term also tends to diverge towards
$H_{c2}$ = 5 T.

Terashima {\it et al.} \cite{TTerashima1} point out that our
$H_{c2}$(onset) = 5 T, determined from the onset of the resistive
transition, is much higher than their $H_{c2}$ = 1.25 T, determined
from the midpoint of the resistive transition. They attribute this
large difference in $H_{c2}$ to the broad resistive transition of
our sample, which indicates inhomogeneity in the sample. Therefore,
they doubt if the $\rho(T) \sim T^{1.5}$ behavior of resistivity at
$H_{c2}$(onset) = 5 T relates to quantum criticality. Their recent
de Haas-van Alphen (dHvA) results \cite{TTerashima2} also do not
support our proposed QCP at $H_{c2}$(onset) = 5 T in KFe$_2$As$_2$.

Recently, we have measured another KFe$_2$As$_2$ single crystal
(S2). As seen in Fig. 1(a), the 10-90\% transition width of S2 is
0.32 K, much smaller than 1.35 K of previous reported sample (S1) in
Ref. [1]. This suggests that S2 is more homogenous than S1. The
sample S2 also has lower residual resistivity $\rho_0$ = 1.49
$\mu$$\Omega$ cm, and higher residual resistivity ratio (RRR)
$\rho$(290 K)/$\rho_0$(3 T) = 265. In Fig. 1(b), $\rho(T)$ of S2
manifests $T^{1.5}$ dependence from $T_c$(onset) up to 11 K in zero
field. From Fig. 1(c), $H_{c2}$(onset) = 3 T is obtained for S2,
where $\rho(T) \sim T^{1.5}$ persists down to 50 mK. When further
increasing the field, the $\rho(T) \sim T^2$ Fermi-liquid behavior
is observed at lowest temperature for S2.

Since $H_{c2}$(onset) of S2 is significantly smaller than that of
S1, we realize that the non-Fermi-liquid behavior of $\rho(T)$ at
$H_{c2}$(onset) does not determine a QCP at $H_{c2}$(onset) for
KFe$_2$As$_2$. In fact, for CeCoIn$_5$, while specific heat data
demonstrated a QCP at the bulk $H_{c2}$ = 5 T, non-Fermi-liquid
$\rho(T) \sim T$ down to lowest temperature was found at higher
field $H$ = 6 T \cite{Bianchi}. We attribute this misfit to the
inhomogeneity of the sample. At the QCP $H_{c2}$ = 5 T, while the
bulk of the CeCoIn$_5$ sample obeys $\rho(T) \sim T$, the rest of
the sample still shows resistive transition, thus one can not
observe $\rho(T) \sim T$ at the QCP. With increasing field, at $H$ =
6 T, the bulk of the sample slightly develops $\rho(T) \sim T^2$
behavior, which balances the remaining resistive drop of the rest
part of the sample, and gives an accidental $\rho(T) \sim T$
behavior. Only for extremely homogeneous sample with nearly zero
resistive transition width, one may not notice this misfit since
$H_{c2}$(onset) is almost equal to the bulk $H_{c2}$. We believe
that this is also the case for KFe$_2$As$_2$, and the QCP, if
exists, may locate at the bulk $H_{c2}$ as in CeCoIn$_5$.

Since the bulk of KFe$_2$As$_2$ have developed Fermi liquid state at
$H$ = 5 T, it is not surprising that dHvA oscillations were observed
in a field range near 5 T \cite{TTerashima2}. For our high-quality
sample S2, we also find that the coefficient $A$ (= 0.0649, 0.0533,
and 0.0508 $\mu$$\Omega$ cm/K$^2$ for 5, 8, and 12 T, respectively)
shows a slower field dependence than that of sample S1. This is
consistent with the near constant effective mass $m^*$ in the field
range $7 < H < 17.65$ T \cite{TTerashima2}, which is far away from
the QCP (if exists) near the bulk $H_{c2} \approx$ 1.25 T.

\begin{figure}
\includegraphics[clip,width=5cm]{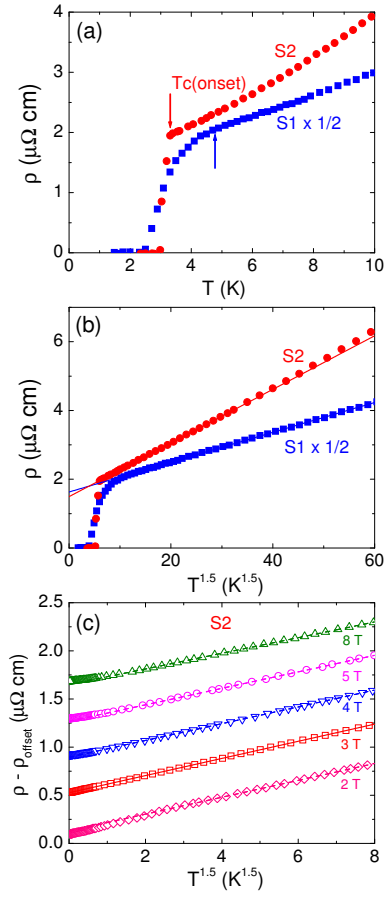}
\caption{(Color online). (a) Low-temperature resistivity of
KFe$_2$As$_2$ single crystals S1 and S2 in zero magnetic field. (b)
The same data in (a) plotted as $\rho$ vs $T^{1.5}$. The solid lines
are fits to $\rho$ = $\rho_{0}$ + $AT^{1.5}$. (c) $\rho$ vs
$T^{1.5}$ for sample S2 in $H$ = 2, 3, 4, 5, and 8 T (data sets are
offset for clarity). The solid line is a fit of the $H$ = 3 T data
between 50 mK and 4 K. The dash lines are guides to the eye for the
deviation from the $T^{1.5}$ dependence.}
\end{figure}

In summary, we agree with Terashima {\it et al.} that
$H_{c2}$(onset) = 5 T is not a QCP for KFe$_2$As$_2$. The
non-Fermi-liquid $\rho(T) \sim T^{1.5}$ at $H_{c2}$(onset) and the
development of a Fermi liquid state with $\rho(T) \sim T^2$ when
further increasing the field only suggest a QCP at the bulk
$H_{c2}$. Bulk measurements, such as specific heat, are needed to
confirm this field-induced QCP at $H_{c2}$ in KFe$_2$As$_2$.

$^*$ E-mail: shiyan$\_$li@fudan.edu.cn

\end{document}